\documentclass[a4paper]{article}

\usepackage{cite}
\usepackage{float}
\usepackage{subfigure}
\usepackage{caption, tabularx}
\usepackage{booktabs}
\usepackage{float}
\usepackage{array}  % 对齐
\usepackage{booktabs} 
\usepackage{multirow}
\usepackage{makecell}

\makeatletter
\newcommand*\bigcdot{\mathpalette\bigcdot@{.5}}
\newcommand*\bigcdot@[2]{\mathbin{\vcenter{\hbox{\scalebox{#2}{$\m@th#1\bullet$}}}}}
\makeatother

\usepackage{INTERSPEECH2020}

\title{Transfer Learning for Improving Singing-voice Detection in \\ Polyphonic Instrumental Music}
\name{Yuanbo Hou$^*$, Frank K. Soong$^\dag$, Jian Luan$^\ddagger$, Shengchen Li$^*$ \thanks{$^*$Work performed as an intern at Microsoft Research Asia.}}
\address{$^*$Beijing University of Posts and Telecommunications\\
$^\dag$Microsoft Research Asia\\
$^\ddagger$Microsoft Search Technology Center Asia, XiaoIce}
\email{\{hyb,shengchen.li\}@bupt.edu.cn, \{frankkps,jianluan\}@microsoft.com}

\begin{document}

\maketitle
\begin{abstract}
  Detecting singing-voice in polyphonic instrumental music is critical to music information retrieval. To train a robust vocal detector, a large dataset marked with \textsl{vocal} or \textsl{non-vocal} label at frame-level is essential. However, frame-level labeling is time-consuming and labor expensive, resulting there is little well-labeled dataset available for singing-voice detection (S-VD). Hence, we propose a data augmentation method for S-VD by transfer learning. In this study, clean speech clips with voice activity endpoints and separate instrumental music clips are artificially added together to simulate polyphonic vocals to train a \textsl{vocal}/\textsl{non-vocal} detector. Due to the different articulation and phonation between speaking and singing, the vocal detector trained with the artificial dataset does not match well with the polyphonic music which is singing vocals together with the instrumental accompaniments. To reduce this mismatch, transfer learning is used to transfer the knowledge learned from the artificial speech-plus-music training set to a small but matched polyphonic dataset, \textsl{i.e.}, singing vocals with accompaniments. By transferring the related knowledge to make up for the lack of well-labeled training data in S-VD, the proposed data augmentation method by transfer learning can improve S-VD performance with an \textsl{F-score} improvement from 89.5\% to 93.2\%.

\end{abstract}

\section{Introduction}

Singing-voice detection (S-VD) is to detect vocal frames of given music clips. Successful detection of singing voice regions in polyphonic music is critical to music information retrieval (MIR) \cite{1} tasks, such as music summarization \cite{2}, retrieval \cite{3}, transcription \cite{4}, genre classification \cite{5}, and vocal separation \cite{6}. 

Recently, deep learning has been applied to S-VD. Deep neural networks \cite{14} are used to estimate an ideal binary spectrogram mask that represents the spectrogram bins in which the vocal is more prominent than the accompaniments. Convolutional neural networks (CNN) have been used to boost the performance in MIR \cite{16}, with an efficient model built on temporal and timbre features. Recurrent neural networks (RNN) are employed to predict time-frequency masks of multiple source signals, then masks are multiplied with the original signal to obtain the desired isolated source \cite{16-1}. Above models can be refined with more accurate frame-level labels, also known as strong labels \cite{16-2}. However, labeling strong label is time-consuming, hence usually datasets have been used with only small number of songs with strong labels in training.

To overcome the limitation of lack of frame-level labeled training data in S-VD, we propose a data augmentation \cite{31, 32} method for S-VD by transfer learning. Transfer learning \cite{17} extracts representations learned from a source task and applies to a similar but different target task. Transfer learning can alleviate the problem of insufficient training data for the target task and tend to generalize the model. Many transfer learning methods \cite{20, 21, 22} related to S-VD use strong labels, and some methods even need clean singing recordings. Datasets with strong labels or clean singing recordings are scarce. However, clean speech corpora and instrumental music datasets are widely available in the Internet, and the endpoints of clean speech can be easily detected. Hence, these clean speech clips and instrumental music clips can be artificially added together to simulate polyphonic vocals for training a vocal detector. To make up for the lack of well-labeled training data in S-VD, this paper proposes to transfer the latent representations of vocal detector in speech-plus-music domain to detect singing voice in polyphonic music domain. Given a source domain $\mathit{D_S}=\{\mathit{X_S},\mathit{f_S}(X)\}$ and source task $\mathit{T_S}$, a target domain $\mathit{D_T}=\{\mathit{X_T},\mathit{f_T}(X)\}$ and target task $\mathit{T_T}$. In this paper, $\mathit{X_S}$ denotes audio clips synthesized by speech clips and instrumental music, $\mathit{T_S}$ is speech activity detection, and $\mathit{f_S}$ is latent representations mapping function learned by the convolutional layers. $\mathit{X_T}$ denotes polyphonic music and $\mathit{T_T}$ is S-VD. Transfer learning \cite{17} aims to improve the learning of the target mapping function $\mathit{f_T}()$ in $\mathit{D_T}$ using the information in $\mathit{T_S}$ and $\mathit{D_S}$.

To investigate the performance of data augmentation by transfer learning in S-VD and explore the possibility of transferring the knowledge from speech to singing voice, the learned representation which retains relevant information of speech clips, will be transferred to S-VD which is a similar but different target task. Although there is difference between speaking and singing, and vocals characteristics may also vary with the change of accompaniments \cite{23}, they still have useful similarities to be exploited. In addition, sharing knowledge of voice between speech clips and the singing voice enable the detector to understand human voice, speech or singing vocal, in a more general and robust form.

The main contributions of this paper are: 1) to overcome the lack of frame-level labeled training data in S-VD, we propose a data augmentation method for S-VD by transfer learning; 2) we investigate the performance of transferring representations learned in speech activity detection to detect singing voice, and find the lower convolutional layers learn more basic local representations which are more effective for detecting vocals in polyphonic music; 3) patterns of convolutional filters are visually analyzed, and the learned knowledge of voice between detectors trained with synthesized audio clips and polyphonic music clips is compared.

The rest of the paper is organized as follows. Section 2 shows the proposed method. Section 3 describes experiments and analyzes results in detail. Section 4 gives the conclusions.

\section{Proposed method}
\label{sec:Proposed method}

\begin{figure}[b]
	\vspace{-0.3cm}  %????????????
	\setlength{\abovecaptionskip}{0.2cm}   %??????????
	\centerline{\includegraphics[width = 0.43  \textwidth]{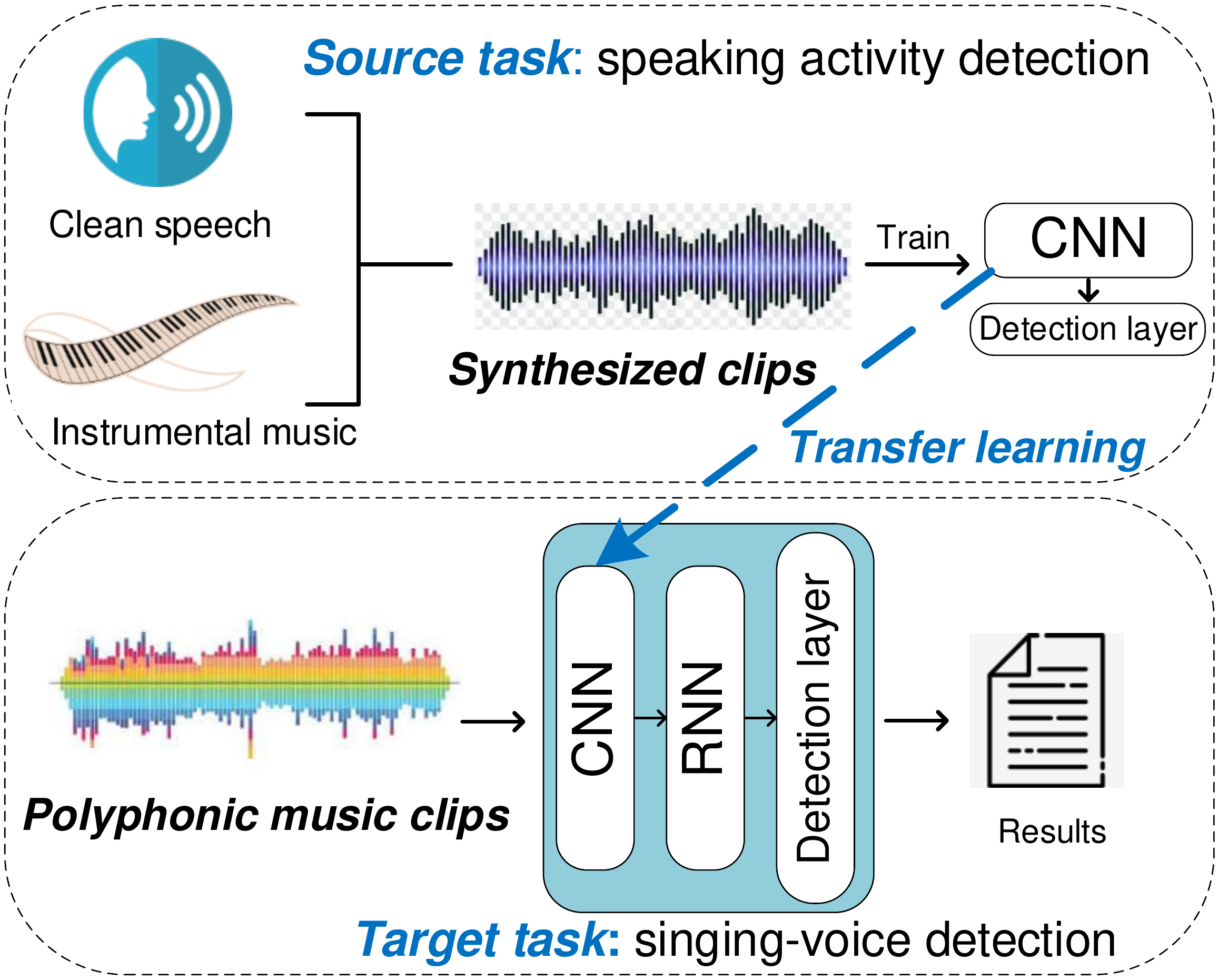}}
	\caption{Framework of the proposed method.}\label{framework}
\end{figure}

The proposed method for S-VD is illustrated in Figure \ref{framework}. To overcome the lack of well-labeled training data in S-VD, transfer learning extracts knowledge of voice from the source task and applies it to the target task to detect singing voice. This is crucial for our task, where the training data for the target task is insufficient to train a good detector model. In the source task, CNN is trained for detecting speech activity frames in synthesized audio clips. The knowledge of voice learned from the large-scale dataset in source task is then transferred to the target task. Due to the different articulation and phonation between speaking and singing  \cite{23}, the target task is more challenging. So a convolutional recurrent neural network (CRNN) is trained with a small set of data collected in the target task to detect the vocal frames.

\label{ssec:framework}

%\vspace{-2.0cm}
\subsection{Source task: speech activity detection}
\begin{figure}[tb]
	%\vspace{-0.0cm}  %????????????
	%\setlength{\abovecaptionskip}{0.05cm}   %??????????
	\setlength{\belowcaptionskip}{-0.45cm}   
	\centerline{\includegraphics[width = 0.48\textwidth]{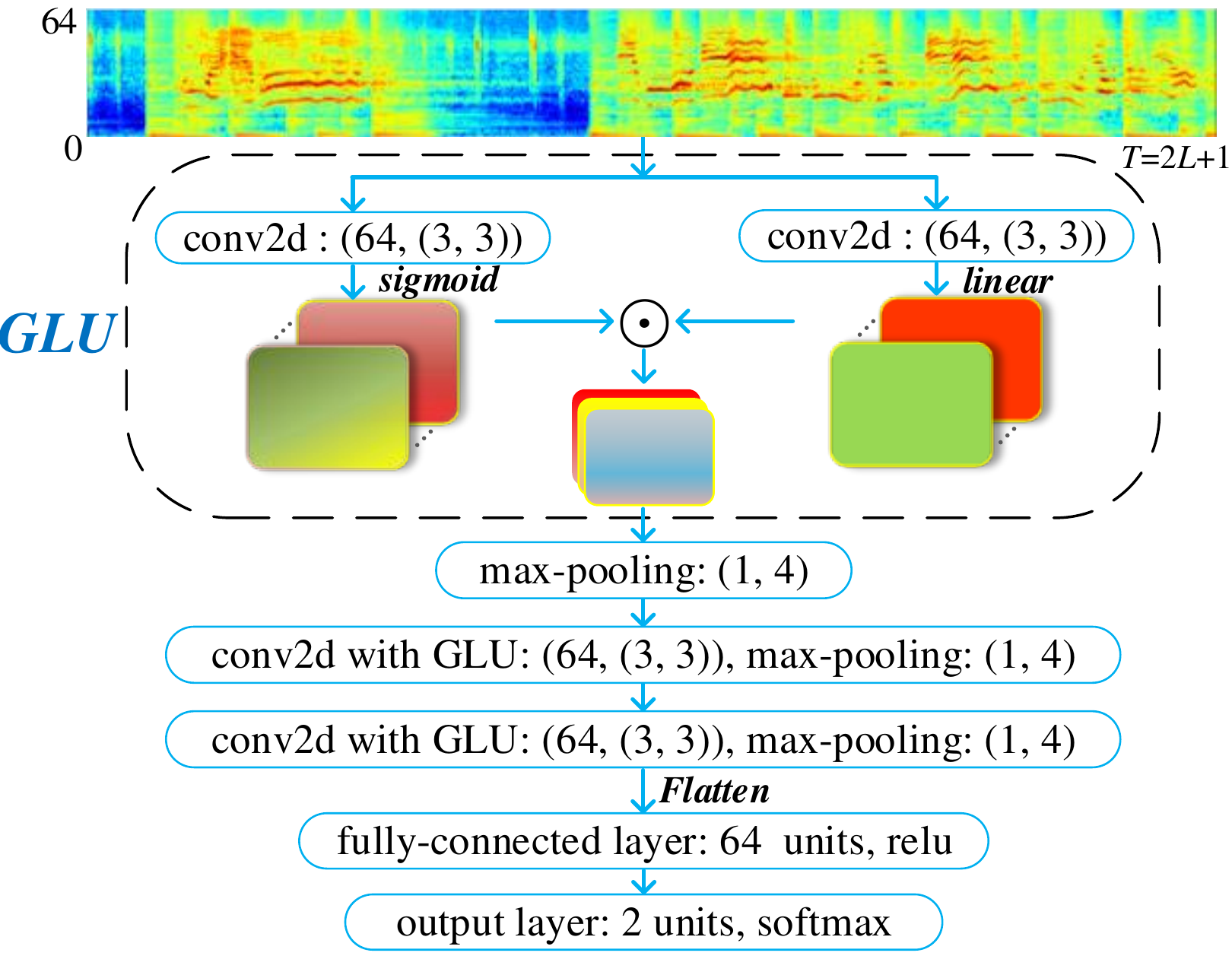}}
	\caption{Details of the CNN architecture in the source task.}\label{cnn}
\end{figure}

The source task is to detect the speech activity endpoints in the synthetic audio clips to learn the representations of voice. For the good performance of CNN in MIR \cite{10, 11}, CNN is used as the detector in the source task. Figure \ref{cnn} shows the details of CNN. The waveforms of synthetic audio clips are converted to log mel spectrogram, which is a 2D representation that approximates human auditory perception. This computationally efficient input has been shown to be effective in MIR tasks such as music classification \cite{24}.

\label{ssec:cnn}

To comprehensively consider the contextual information of audio, the input of CNN is a moving data block, consisting of the preceding \textsl{L} frames and the succeeding \textsl{L} frames of the current frame, the shift between succeeding blocks is one frame. Each block contains \textsl{(2L+1)} frames. \textsl{L} determines the range of contexts visible in the model at every frame.

The detector consists of a series of convolutional and pooling layers. To preserve the time resolution of the input, pooling is applied to the frequency axis only. As shown in Figure \ref{cnn}, where (64, (3, 3)) corresponds to (convolutional filters, (receptive field in time, frequency)). Pooling layer is specified by (pooling length in time, frequency). In addition, to reduce the gradient vanishing problem in deep networks training, gated linear units (GLUs) \cite{25} are used in convolutional layers. They provide a linear path for gradient propagation while keeping nonlinear capabilities through the sigmoid operation. Given \textsl{W} and \textsl{V} as convolutional filters, \textsl{b} and \textsl{c} as biases, \textsl{X} as the input features or the feature maps of interval layers and $\sigma$ as sigmoid function, GLUs are defined as:
\begin{equation}
\setlength{\abovedisplayskip}{3pt}
\setlength{\belowdisplayskip}{3pt}
Y=(W \ast X + b)\odot \sigma (V \ast X +c)
\end{equation}
where the symbol $\odot$ is the element-wise product and $\ast$ is the convolution operator. By weighting time-frequency units according to their unique time positions, GLUs can help network attend to voice and ignore unrelated accompaniments.

The source task aims to detect whether there is speech in a frame, which is a binary classification task. If sigmoid function with one unit is used in the last layer of the CNN, thresholds are needed to determine the label of each frame. To avoid the impact of thresholds on detection results, softmax function with two output units are used in the last layer. The label corresponding to the larger output probability is used as the final label of each frame. 

%\vspace{-0.2cm}
\subsection{Target task: singing-voice detection}
When the detection aims at polyphonic songs, relying on the CNN trained on the artificial synthesized audio clips may be inadequate, because both articulation and phonation between speech and singing are different \cite{23}. In addition, the vocals in polyphonic music will change together with the accompaniments. It is known that singing voice evolves in songs, which can bring more variation to the vocal representations. 

Compared with the source task based on synthesized data, the target task is more challenging. Vocals, which change together with the accompaniments, are difficult to detect in polyphonic music, so a recurrent layer is added to the CNN to capture the long-term temporal contextual information of audio signal. In the target task, the detector is a convolutional recurrent neural network (CRNN), which adds a recurrent layer after the last convolutional layer of the CNN in Figure \ref{cnn}. The rest of the CRNN is consistent with the CNN in Figure \ref{cnn}. 

There are two modes for transferring knowledge from the source task to the target task depending on whether the transferred parameters are updated during the training phase in the target task. In this paper, a comparative study is conducted to investigate the effects of two modes on the proposed system.

%\vspace{-0.2cm}
\subsection{Visualizing the patterns of convolutional filters}

It is difficult to display or measure the knowledge in speech and sing voice directly. Fortunately, convolutional layers in the model can extract the features of the input data, which are indirect representations of the knowledge contained in the speech and singing voice. To intuitively inspect the differences of knowledge in speech and singing voice, the gradient ascent \cite{27} is used to show the patterns learned from the data by convolutional filters. Given $\mathit{X}$ is a blank input image, $\mathit{x}$ is the point in $\mathit{X}$, $\mathit{\eta}$ is learning rate and $\mathit{a_{ij}(x)}$ is the output of the filter at $\mathit{(i,j)}$ after convolution. The pattern of the filter can be calculated by:

\begin{equation}
\setlength{\abovedisplayskip}{3pt}
\setlength{\belowdisplayskip}{3pt}
\mathit{X}=\mathit{X} + \mathit{\eta}{ \partial a_{ij}(x)}/{\partial x}
\end{equation}

The visualization method applies gradient descent to the value of the input image of a convolutional layer so as to maximize the response of a specific filter. Repeat this step many times, the resulting image will be one that the chosen filter is maximally responsive to, \textsl{i.e.} the pattern of the filter.

\section{Experiments and results}
%\vspace{-0.3cm}
\subsection{Dataset and Experiments Setup}
For the source task, artificially synthesized audio clips are required to train the CNN, which is able to learn the spectral and temporal features of speech signal. For this reason, a private clean speech corpus from Microsoft XiaoIce group with 100 speakers, each speaker recorded about 20 minutes of speech, in total for about 34.5 hours, was artificially added together with an instrumental music dataset at signal-to-noise ratio of 0 dB to simulate polyphonic music clips. The endpoints of voice in the clean speech are detected, hence the frame-level label of the synthesized polyphonic audio clips are obtained, accordingly. 

For the target task, the dataset consisting of 120 polyphonic songs is divided into training and validation sets. The test set consists of another 60 polyphonic songs. Each song in the target task is about 4 minutes long and there is no intersection of singers in training, validation and test sets. These songs are annotated with frame-level \textsl{on}/\textsl{off} labels as the ground truth representing the singing voice is on or not in each audio frame. More details, source codes and samples, please see here\protect\footnote{https://github.com/moses1994/singing-voice-detection}.

In training, log mel spectrogram is extracted using STFT with Hamming window length of 40 ms, which has sufficient time and frequency resolution. An overlap of 50\% between two adjacent windows is used to smooth the spectrograms. Then 64 mel filter banks are applied. Dropout and normalization are used to prevent over-fitting. Both the source and target tasks are binary classification tasks, hence Adam optimizer \cite{adam} is used to minimize the binary cross entropy.

Given the frame-wise detection results for each frame, we can calculate precision (\textsl{P}), recall (\textsl{R}) and \textsl{F-score} (\textsl{F}) of the detection performance. They are defined as:
\begin{equation}
\setlength{\abovedisplayskip}{2.5pt}
\setlength{\belowdisplayskip}{2.5pt}
P=\frac{N_{tp}}{N_{tp}+N_{fp}},    R=\frac{N_{tp}}{N_{tp}+N_{fn}},    F=\frac{2P\cdot R}{P+R}
\end{equation}
where $N_{tp}$, $N_{fp}$ and $N_{fn}$ are the numbers of true positives, false positives and false negatives, respectively. Higher \textsl{P}, \textsl{R} and \textsl{F} indicate a better performance \cite{29}.

%\vspace{-0.35cm}
\subsection{Results and analysis}

\begin{figure}[t]
	%\vspace{-0.3cm}  %????????????
	\setlength{\abovecaptionskip}{0.2cm}   %??????????
	\setlength{\belowcaptionskip}{-0.5cm}   
	\centerline{\includegraphics[width = 0.5  \textwidth]{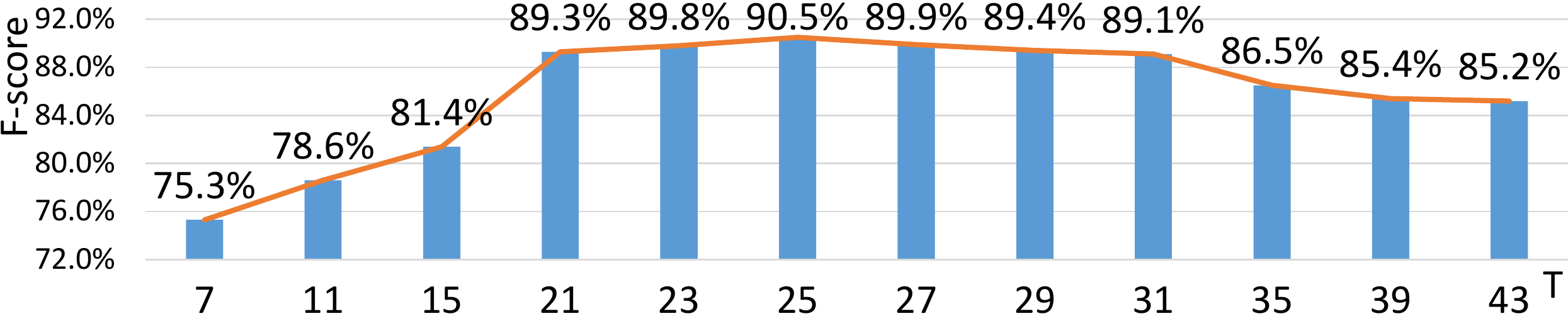}}
	\caption{Results of different input lengths in the source task.}\medskip\label{l}
\end{figure}

To consider the long-term contextual information of the audio clips, the input of CNN is a block totaling (\textsl{2L+1}) frames. Figure \ref{l} shows the results of CNN trained with blocks of different lengths, on the \textsl{x}-axis is different values of \textsl{T} frames, \textsl{i.e.} (\textsl{2L+1}) frames, and on the \textsl{y}-axis is \textsl{F-score}. The comparison in Figure \ref{l} reveals that performance of detector does not improve monotonically with increased length of input block, and setting \textsl{T}=25 achieved a good trade-off between \textsl{F-score} and computational complexity. Consequently, this value is used for all later experiments.

\label{ssec:l}

%\hline 表示画出一整条从左至右横线
%\cline{2-3} 表示画出一条在第2栏位到第3栏位的横线段，其他栏位将不会有横线段
%跨列（将两列合并为一列）第一个参数指明跨几列  第二个参数指明内容居中并在左右两边画上直线 最后一个参数是表格内容
\begin{table}[b]\footnotesize
	% 表格标题的距离 above设置标题上面的距离，below设置标题下面的距离
	\setlength{\abovecaptionskip}{0.2cm}   %表格和标题之间的距离
	\setlength{\belowcaptionskip}{-0.25cm}   %???????????????????
	\renewcommand\tabcolsep{0.5pt} 
	%对于一个大的表格，可用 \setlength{\tabcolsep}{1pt}来减少表格的列间距离；也可用\resizebox{!}{5cm}{\begin{tabular} ... \end{tabular}}把整个表格当作一个图形
	\centering
	\caption{The results of two different transfer modes.}
	%\begin{tabular}{c|c|c|c|c}
	\begin{tabular}{p{2.5cm}<{\centering}|p{1.2cm}<{\centering}|p{1.4cm}<{\centering}|p{1.2cm}<{\centering}|p{1.4cm}<{\centering}}
		%1.5cm是Entire CNN 调整距离
		%p{1.345cm}<{\centering} 调整后数据居中
		\hline
		\multirow{2}{*}{\makecell[c]{\textbf{\textsl{Transferred layer}}}}  & \multicolumn{2}{c|}{\textbf{\textsl{{Fixed}}}} & \multicolumn{2}{c}{\textbf{\textsl{Fine-tuning}}}  \\
		\cline{2-5}
		& \textsl{F-score} & \textsl{N.params} & \textsl{F-score} & \textsl{N.params} \\
		\hline 		
		%\hline
		\textsl{$L_1$} & 91.9\% & 20.58\textsl{K} & \textbf{93.2\%} & 20.72\textsl{K} \\
		%\hline
		\textsl{$L_2$} & 91.7\% & 13.33\textsl{K}  & 92.0\% & 20.72\textsl{K} \\
		%\hline
		\textsl{$L_3$} & 91.1\% & 13.33\textsl{K}  & 91.7\% & 20.72\textsl{K} \\
		\textsl{$L_{all}$} & 82.6\% & 5.79\textsl{K} & 92.3\% & 20.72\textsl{K} \\
		\hline
	\end{tabular}
	\label{tab:two_modes}
\end{table}

Given \textsl{$L_1$}, \textsl{$L_2$} and \textsl{$L_3$} denote the first, the second and the third convolutional layer with GLUs, \textsl{$L_{all}$} denotes all convolutional layers. In transfer learning, the \textsl{$L_i$} in CRNN in the target task will accept the learned parameters of \textsl{$L_i$} in CNN in the source task. In \textsl{Fixed} mode, the parameters of \textsl{$L_i$} in CRNN will no longer be updated during the backpropagation, other layers of CRNN are trained normally. In \textsl{Fine-tuning} mode, the \textsl{$L_i$} in CRNN will continue to adapt its parameters with the target dataset. Due to the limitation of space, \textsl{F-score} of two modes on the test dataset of target task, and the number of trainable parameters (\textsl{N.params}) are shown in Table \ref{tab:two_modes}.

\begin{figure}[t]
%\vspace{-0.27cm}
\setlength{\abovecaptionskip}{-0cm}   
\setlength{\belowcaptionskip}{-0.8cm}
\centering\includegraphics[width = 0.35\textwidth]{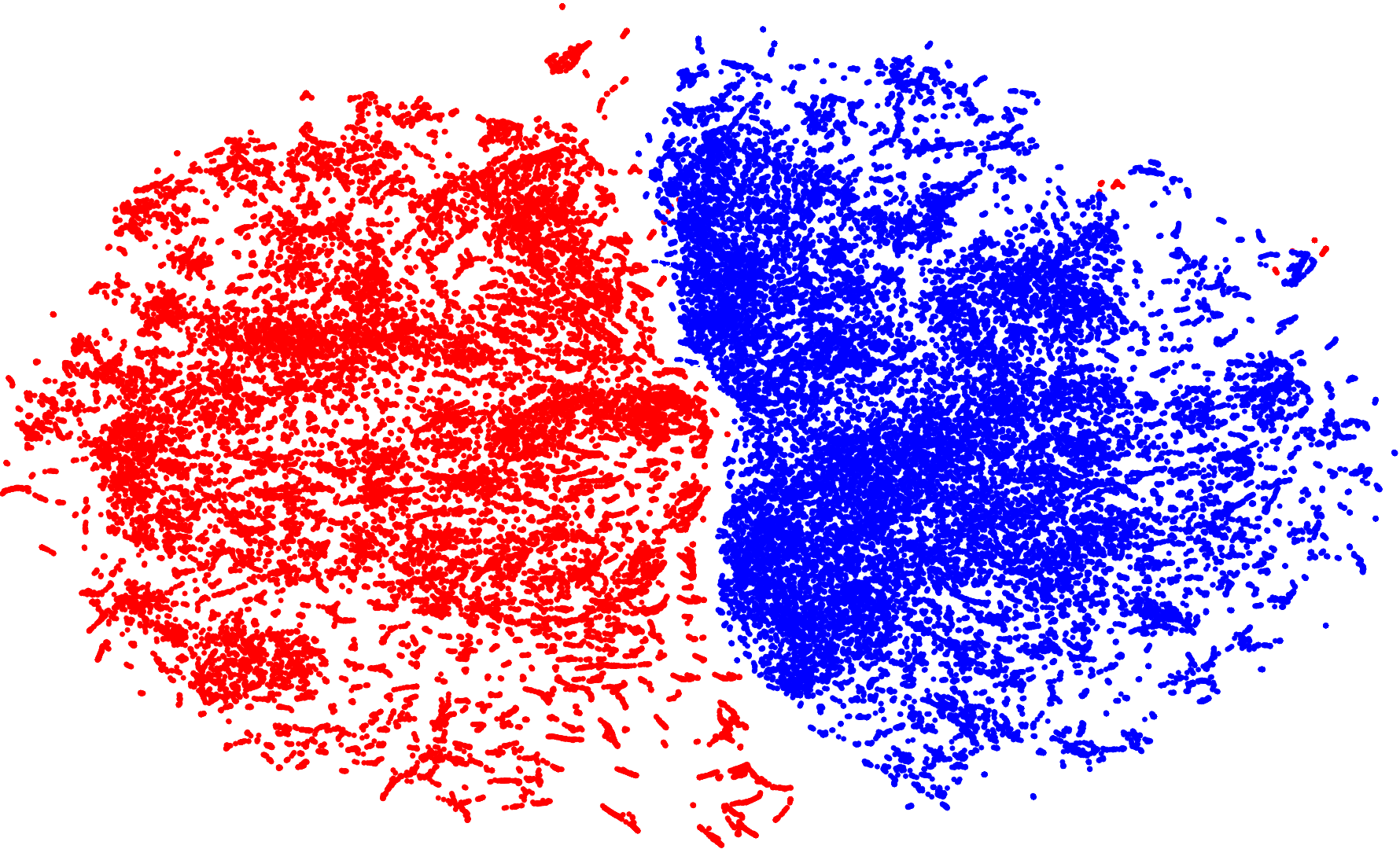}
%\caption{Clustering result.}\medskip\label{tsne}
\caption{Visualization of features distribution using \textsl{t-SNE} \cite{tsne}, the red points and blue points denote singing voice samples in the target task and synthesized polyphonic audio samples in the source task, respectively.}\medskip\label{tsne}
\end{figure}

As shown in Table \ref{tab:two_modes}, the performance of transferring the all convolutional layers of the CNN in the source task and freeze them yields the worst result. However, transferring \textsl{$L_1$} with fine-tuning yields the best result. Transferring the knowledge of \textsl{$L_2$} or \textsl{$L_3$} does not perform as well as \textsl{$L_1$}. This may due to lower level convolutional layers may contain more generic features (e.g. edge or frequency detectors) that are useful for both source and target tasks. They learn the basic and local features of voice, but high level convolutional layers may become more irrelevant in learning some high level representations. The singing voice in the target task is more complex than the speech in the source task, because the singing voice will change with the polyphonic accompaniments. Hence, the high level representations of voice learned by the higher convolutional layers from speech may not match the target task, resulting transferring this knowledge does little in helping the target task. To show the difference between source domain and target domain more intuitively, Figure \ref{tsne} shows the results of high-dimensional acoustic features clustering of synthesized polyphonic audio samples and singing voice samples in polyphonic music by \textsl{t-SNE} \cite{tsne}. It can be seen from the Figure \ref{tsne} that the features of the synthesized polyphonic audio samples in the source task are clearly separated from the features of the actual singing voice samples in the target task after high-dimensional clustering. Therefore, the synthesized polyphonic audio samples cannot completely simulate the characteristics of the singing voice in polyphonic instrumental music, which leads to the fact that the knowledge learned by the vocal detector from the source task cannot be fully applied to the target task.

\begin{figure}[hbt]
	%\vspace{-0.27cm}
	%\setlength{\abovecaptionskip}{0cm}   
	\setlength{\belowcaptionskip}{-0.4cm}
	\centering\includegraphics[width = 0.46\textwidth]{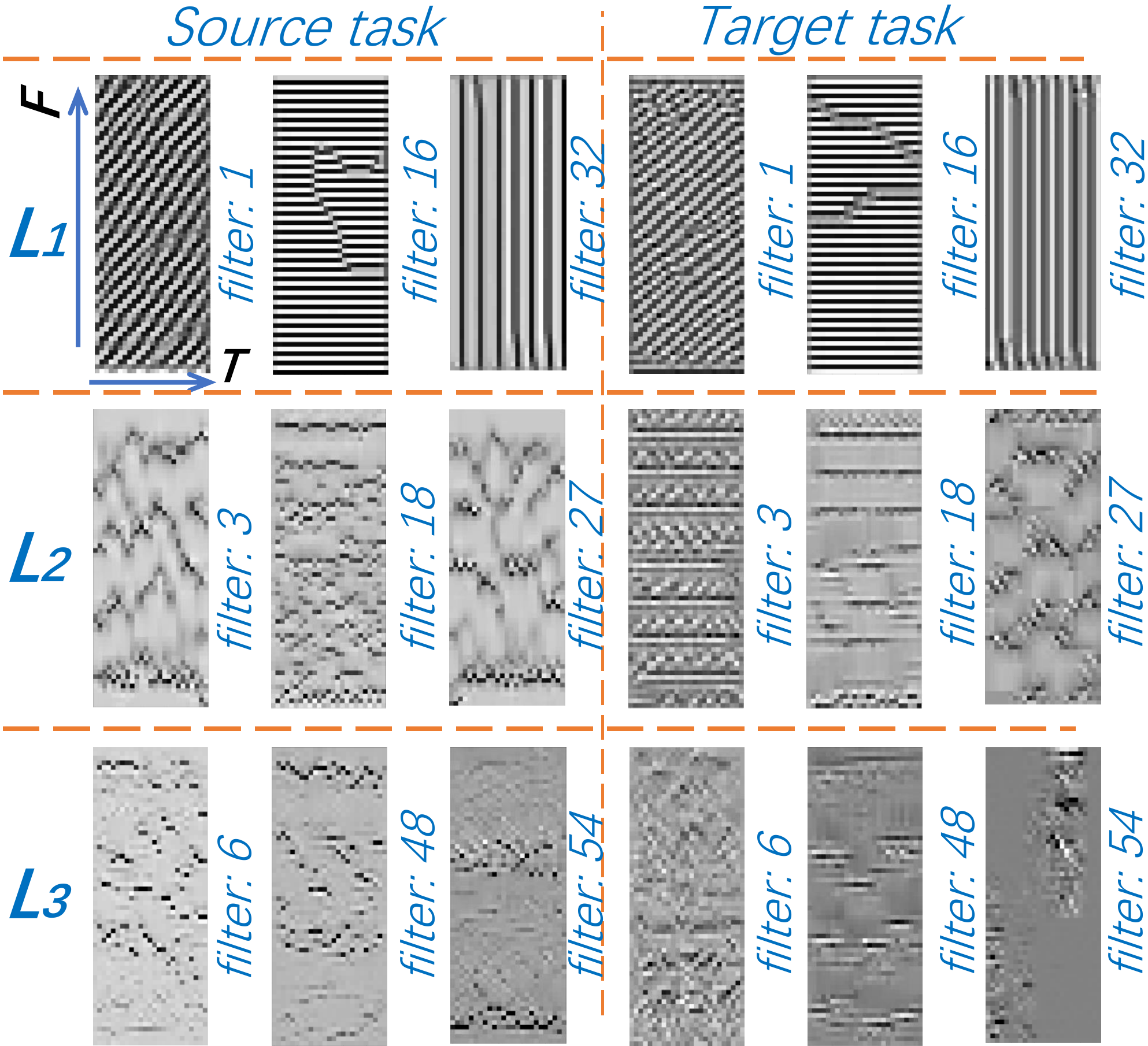}
	%\caption{Clustering result.}\medskip\label{tsne}
	\caption{Patterns of different filters in \textsl{$L_i$}, for each subgraph, the x-axis is time (\textsl{$T$}) and the y-axis is frequency (\textsl{$F$}).  
	}\medskip\label{pattern}
\end{figure}

To gain deeper insights of the knowledge in the source and target tasks, we visualized the learned patterns of filters in convolutional layers with GLUs. Due to the limitation of space, patterns which are randomly selected from different filters, is shown in Figure \ref{pattern}. Please see here$^1$ for more details. In Figure \ref{pattern}, for the same model in a task, \textsl{$L_1$} learns more obvious basic local features of the input spectrogram than \textsl{$L_2$} and \textsl{$L_3$}. For different models in the two tasks, compared with the learned patterns of \textsl{$L_2$} and \textsl{$L_3$}, the patterns of \textsl{$L_1$} in the two task are more similar. This may be the reason why transferred \textsl{$L_1$} performs best in Table \ref{tab:two_modes}. Since the local representations of voice learned in the source task and target task are relatively similar, transferring this knowledge to target domain can help the model obtain a more general and robust vocal detection. For the \textsl{$L_2$} and \textsl{$L_3$}, the high level representations they learned from different domain is quite different, hence transferring this knowledge provides little help to the target task.

\begin{figure}[t]
	\setlength{\belowcaptionskip}{-0.65cm}   %???????????????????
	\centerline{\includegraphics[width = 0.50\textwidth, height=1.5cm]{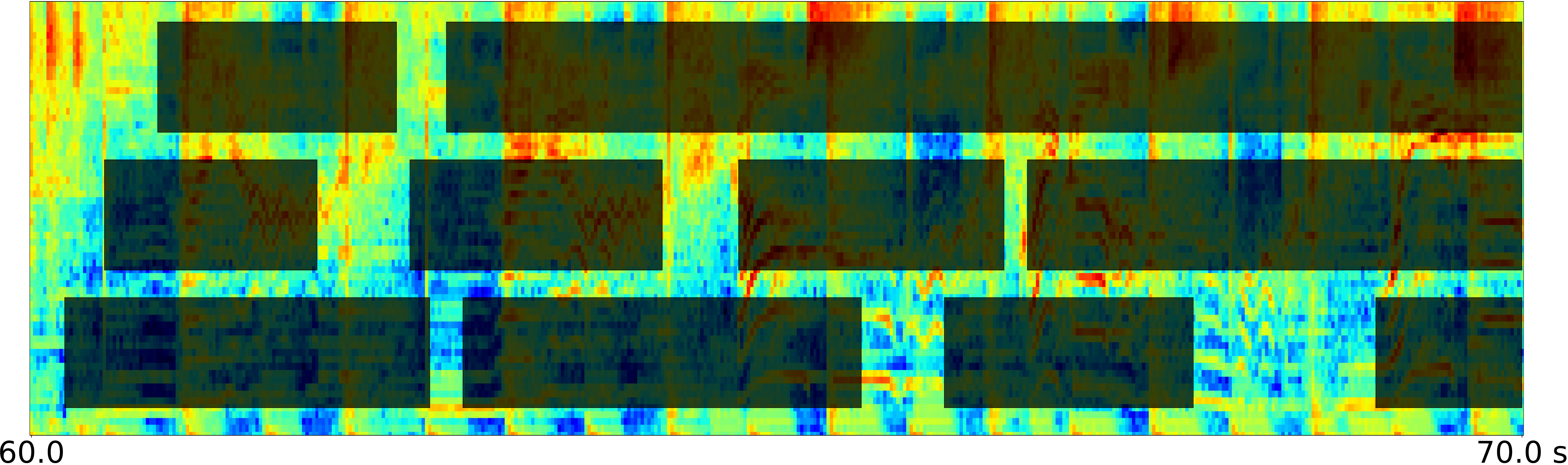}}
	\caption{From top to bottom, they are visualization of the ground truth, the results of proposed method and baseline, respectively. Shaded parts indicate singing voice activity.}\medskip\label{below}
\end{figure}

When the optimal transfer mode is determined, the detection results on the test set in the target task are shown in Table \ref{tab:10songs}. The baseline is a deep CNN architecture with 3-by-3 2D convolution layers \cite{28} trained directly with the dataset in the target task. And \cite{28} implies that CNN may benefit from looking at a varying range of time and frequency to learn vocal-specific characteristics, such as timbre \cite{timbre}. For most polyphonic songs 
in Table \ref{tab:10songs}, the results of the proposed data augmentation method by transfer learning have better \textsl{F-score} higher than the baseline. A very robust sample of detection results is shown in Figure \ref{below}.

The singing-voice detector trained by the transfer learning was also tested on MUSDB18 \cite{musdb18} to compare the performance on the publicly available music dataset. MUSDB18 contains 150 tracks ($\sim$10h duration) of different styles, the 150 tracks are split into 100 tracks for training, and 50 for testing. The detection results on the test set in MUSDB18 are shown in Table \ref{tab:himusdb}. In addition to the precision, the model trained by transfer learning in this paper is better than the baseline in recall and \textsl{F-score}. The reason may be that the training data in this paper has more types of samples, and the model can learn more different information in the process of transfer learning.

\begin{table}[h]\footnotesize
%\vspace{-0.2cm}
	% 表格标题的距离 above设置标题上面的距离，below设置标题下面的距离
	\setlength{\abovecaptionskip}{0.2cm}   %表格和标题之间的距离
	\setlength{\belowcaptionskip}{-0cm}   %???????????????????
	\renewcommand\tabcolsep{0.5pt} 
	%对于一个大的表格，可用 \setlength{\tabcolsep}{1pt}来减少表格的列间距离；也可用\resizebox{!}{5cm}{\begin{tabular} ... \end{tabular}}把整个表格当作一个图形
	\centering
	\caption{The detection results on the test set in the target task.}
	\begin{tabular}
		{ p{1.6cm}<{\centering}|
			p{0.80cm}<{\centering}
			p{0.80cm}<{\centering}|
			p{0.80cm}<{\centering}
			p{0.80cm}<{\centering}
			p{0.78cm}<{\centering}|
			p{0.80cm}<{\centering}
			p{0.80cm}<{\centering}
			p{0.80cm}<{\centering}}
		\hline
		\multirow{2}{*}{\makecell[c]{ \textbf{\textsl{Polyphonic}} \\ \textbf{\textsl{song}}}}  
		& \multicolumn{2}{c|}{\textbf{\textsl{Frames}}}
		& \multicolumn{3}{c|}{\textbf{\textsl{{Baseline \cite{28}}}}}
		& \multicolumn{3}{c}{\textbf{\textsl{{Transfer learning}}}}  \\
		\cline{2-9}
		& \textsl{off} & \textsl{on} & \textsl{P (\%)} & \textsl{R (\%)} & \textsl{F (\%)} & \textsl{P (\%)} & \textsl{R (\%)} & \textsl{F (\%)} \\
		\hline 
		\textsl{No.1}  & 2938  & 5384 & 82.8  &  85.4 & 84.1  & 92.9 & 97.3 & \textbf{95.1}  \\
		\textsl{No.2}  & 4166 & 7476 &  83.0 & 90.0  & 86.3  & 89.6 & 98.6 & \textbf{93.9} \\
		\textsl{No.3}  & 4945 & 5754 &  86.5  &  91.6  & 89.0  & 89.3 & 96.5 & \textbf{92.8}  \\		
		\textsl{No.4} & 3390 & 6098  & 79.6 & 91.4 & 85.1 & 84.1 & 91.8 & \textbf{87.8} \\
		\textsl{No.5} & 5844  & 8366  & 96.4  &  93.2  &  89.7  & 88.4 & 92.9  & \textbf{90.6}  \\
		\textsl{No.6}  & 2744 & 4793 &  84.5  &  92.3  &  88.2  & 86.5 & 91.7 & \textbf{89.1}  \\
		\textsl{No.7}  & 6423 & 2911 &  89.5 & 94.4 & \textbf{91.9} & 86.7 & 93.7 &  90.1 \\
		\textsl{No.8}  & 1475 & 4561 & 90.2  &  94.3  &  92.2  & 91.0 & 97.8 &  \textbf{94.2} \\		
		\textsl{No.9}  & 2458  & 9922 &  66.6  & 89.9  & 76.5  & 70.8 & 91.4 & \textbf{79.8}  \\
		\hline
		\multicolumn{1}{c}{\textsl{$\bigcdot\bigcdot\bigcdot$}}  
		& {$\bigcdot\bigcdot\bigcdot$} & 
		\multicolumn{1}{c}{\textsl{$\bigcdot\bigcdot\bigcdot$}} 
		& {$\bigcdot\bigcdot\bigcdot$}  &  {$\bigcdot\bigcdot\bigcdot$} 
		& \multicolumn{1}{c}{\textsl{$\bigcdot\bigcdot\bigcdot$}}
		& {$\bigcdot\bigcdot\bigcdot$} 
		& {$\bigcdot\bigcdot\bigcdot$} 
		& {$\bigcdot\bigcdot\bigcdot$}  \\
		\hline
		\textsl{No.60}  & 3218 & 7220 &  96.5  & 95.7  & 96.1  & 95.8 & 97.7 & \textbf{96.8}  \\
		\hline
		\multicolumn{3}{c|}{\textsl{\textbf{Overall}}}  &  86.1  & 93.2  & 89.5  & 90.1 & 96.0 & \textbf{93.2}  \\
		\hline
	\end{tabular}
	\label{tab:10songs}
\end{table}

\begin{table}[h]
\vspace{-0.4cm}
\setlength{\abovecaptionskip}{0.2cm}   %表格和标题之间的距离
\setlength{\belowcaptionskip}{-0cm}  
\centering
\caption{The detection results on the test set in MUSDB18 \cite{musdb18}}
\begin{tabular}
{ 		    p{0.80cm}<{\centering}
			p{0.80cm}<{\centering}
			p{0.80cm}<{\centering}|
			p{0.80cm}<{\centering}
			p{0.80cm}<{\centering}
			p{0.80cm}<{\centering}}
\hline
\multicolumn{3}{c|}{\textbf{\textsl{{Baseline \cite{28}}}}}
& \multicolumn{3}{c}{\textbf{\textsl{{Transfer learning}}}}  \\
\hline
\textsl{P (\%)} & \textsl{R (\%)} & \textsl{F (\%)} & \textsl{P (\%)} & \textsl{R (\%)} & \textsl{F (\%)} \\
\hline
96.83  & 81.64  & 88.61  & 92.98 & 96.57 & 94.74  \\
\hline
\end{tabular}
\label{tab:himusdb}
\end{table}

%\vspace{-0.3cm}
\section{Conclusions}
To overcome the limitation of insufficient frame-level labeled training data in S-VD, this paper proposes a data augmentation method for S-VD by transfer learning. Due to the shortage of well-labeled polyphonic music data, a training set of clean speech and instrumental music are added together to construct the basic training dataset. The knowledge learned from the artificial training set is then transferred to a small but more matched dataset of singing vocals with instrumental accompaniments, by adapting the corresponding detector parameters to make a better singing voice detector.

By analyzing the patterns of filters, we found the patterns learned from the source task does not match well with target task. This mismatch can be reduced by fine-tuning the convolutional filters parameters at the lower layers of the model. By transferring the related knowledge to make up for the lack of well-labeled training data in S-VD, the proposed data augmentation method by transfer learning can improve S-VD performance with an F-score improvement from 89.5\% to 93.2\%.

\vfill\pagebreak

\bibliographystyle{IEEEtran}

\bibliography{mybib}

% \begin{thebibliography}{9}
% \bibitem[1]{Davis80-COP}
%   S.\ B.\ Davis and P.\ Mermelstein,
%   ``Comparison of parametric representation for monosyllabic word recognition in continuously spoken sentences,''
%   \textit{IEEE Transactions on Acoustics, Speech and Signal Processing}, vol.~28, no.~4, pp.~357--366, 1980.
% \bibitem[2]{Rabiner89-ATO}
%   L.\ R.\ Rabiner,
%   ``A tutorial on hidden Markov models and selected applications in speech recognition,''
%   \textit{Proceedings of the IEEE}, vol.~77, no.~2, pp.~257-286, 1989.
% \bibitem[3]{Hastie09-TEO}
%   T.\ Hastie, R.\ Tibshirani, and J.\ Friedman,
%   \textit{The Elements of Statistical Learning -- Data Mining, Inference, and Prediction}.
%   New York: Springer, 2009.
% \bibitem[4]{YourName17-XXX}
%   F.\ Lastname1, F.\ Lastname2, and F.\ Lastname3,
%   ``Title of your INTERSPEECH 2020 publication,''
%   in \textit{Interspeech 2020 -- 20\textsuperscript{th} Annual Conference of the International Speech Communication Association, September 15-19, Graz, Austria, Proceedings, Proceedings}, 2020, pp.~100--104.
% \end{thebibliography}

\end{document}